\def\year{2020}\relax
\documentclass[letterpaper]{article} 
\usepackage{aaai20}  
\usepackage{times}  
\usepackage{helvet} 
\usepackage{courier}  
\usepackage[hyphens]{url}  
\usepackage{graphicx} 
\usepackage{graphicx}  
\usepackage{xspace}
\usepackage{amsmath}
\usepackage{comment}
\usepackage{array}
\usepackage{color}
\usepackage{mathtools}
\usepackage{dsfont}
\usepackage{booktabs} 

\newcommand{\kenneth}[1]{{\small\color{blue}{\bf\xspace#1 -Kenneth}}}

\newcommand{\patricia}[1]{{\small\color{red}{\bf\xspace#1 -Patricia}}}

\title{On Using Chatbots to Promote Smoking Cessation Among\\Adolescents of Low Socioeconomic Status}
\author{
Patricia Simon,~~Suchitra Krishnan-Sarin\\
Department of Psychiatry\\
Yale University School of Medicine\\
New Haven, CT 06511, USA\\
\texttt{\{p.simon,suchitra.krishnan-sarin\}@yale.edu}\\
\And
Ting-Hao (Kenneth) Huang\\
College of Information Sciences and Technology\\
Pennsylvania State University\\
University Park, PA, USA\\
\texttt{txh710@psu.edu}\\
}

\usepackage{enumitem}

\newcommand{\system}{MyQuitBot\xspace}

\newcommand{\chorus}{Chorus\xspace}

\newcommand{\ev}{Evorus\xspace}

\begin{document}
\maketitle

\begin{abstract}

Reducing youth tobacco use is critical for improving child health since tobacco use is associated with respiratory problems, and nicotine may interfere with healthy brain development. While tobacco regulation has contributed to declines in cigarette use among youth, these declines have occurred more quickly for youth of high socioeconomic status (SES) compared to youth of low SES. A major barrier to smoking cessation for adolescents of low SES is coordination of access and transportation to in-person treatment sessions. Low-SES youth may have family obligations that limit their ability to access in-person treatment. At the same time, mobile use among adolescents is high: 85\% have smartphones. Additionally, adolescents engage in texting at high rates, suggesting that they are well-suited for mobile instant messaging interventions. Mobile interventions have shown promise for youth, but their use remains low. Thus, more research is needed to develop effective and engaging mobile interventions to increase quit rates. In this paper, we provide a brief review of approaches to adolescent smoking cessation and describe the promise of chatbots for smoking cessation.

\end{abstract}

\section{Introduction}
Adolescent smoking is a significant public health concern, especially among individuals of low socioeconomic status (SES). Smoking is a leading cause of preventable morbidity and mortality~\cite{nih25}, with low SES being associated with disproportionate experience of morbidity and mortality from smoking-related disease~\cite{nih3}. The fact that 90\% of adult smokers start in adolescence is a concern.
Although there has been substantial research examining the association between SES and adult tobacco use, there is still much to be explored among adolescents, especially regarding treatment. Although youth smoking rates are declining, nearly 10\% of high-school-aged youth have smoked cigarettes~\cite{nih26}. Youth smoking is of concern because early nicotine exposure may interfere with healthy brain development~\cite{nih1}. Moreover, while tobacco regulation has contributed to declines in cigarette use among youth, these declines have occurred more quickly for high-SES adolescents than low-SES youth~\cite{nih3,nih4}. Pilot data we collected in Connecticut shows that lower SES was associated with increased odds of cigarette smoking when controlling for race, age, and gender (OR 1.8; 95\% CI 1.3, 2.3)~\cite{simon2015socioeconomic}. 
Encouragingly, more than half of adolescent and young adult smokers have intentions to quit~\cite{nih5}. Unfortunately, only 4-6\% of unassisted quit attempts are successful~\cite{nih6}, pointing to the need for professionally developed support methods for smoking cessation.

Smoking cessation interventions that include \textbf{motivational interviewing (MI)} and \textbf{cognitive behavioral therapy (CBT)} elements may be efficacious~\cite{nih27,nih28}. In our review of smoking cessation programs for adolescents, we observed the highest self-reported rate of abstinence (47.5\%) for a program that included MI and CBT~\cite{nih28}. However, these treatments and interventions are not always accessible to adolescents of low SES, who may have family obligations ({\em e.g.}, babysitting or employment to contribute to the household) that limit their ability to access in-person treatment.

Meanwhile, mobile use among adolescents is high: 85\% have smartphones~\cite{nih31}. Additionally, 47\% of youth send more than 50 texts per day, suggesting that they are well-suited for mobile instant messaging interventions~\cite{nih8}. Mobile platforms ({\em i.e.}, phones and tablets) have the potential to deliver scalable, effective, accessible interventions, with high fidelity at a low cost for adolescents of low SES. 
While mobile smoking cessation interventions have shown promise for youth~\cite{nih9}, adoption remains low~\cite{nih10}. Nearly 50\% of treatment-seeking adolescent smokers are interested in mobile smoking cessation programs, and only a third have used their mobile device or the internet for smoking cessation~\cite{nih10}. Mobile interventions that can simulate human conversation, such as chatbots, may attract more youth to pursue smoking cessation.

\section{Chatbots for Smoking Cessation}

Conversational agents capitalize on recent advances in artificial intelligence to extend our ability to provide more human-like mobile interventions. Conversational agents may be deployed in a variety of formats. They may be embodied ({\em i.e.}, having a 2-D or 3-D human-like physical representation) or non-embodied ({\em i.e.}, using text only). Conversational agents may also vary in the type of user input they support, processing speech, text, and nonverbal behaviors. 
Given adolescents' frequent use of texting platforms, we argue that a text-based chatbot is a suitable tool for long-term mobile intervention. One classic example is Woebot, a chatbot developed to promote mental health~\cite{molteni_2018}. Woebot debuted as the first non-embodied chatbot to provide CBT for anxiety and depression. Users can initiate conversations with the chatbot, as they would with a friend on Facebook messenger. Woebot also checks in with users periodically during the day to ask about their mood and suggest coping strategies. Preliminary findings show significant reductions in depression among young adults after two weeks~\cite{nih33}.

Despite recent advances in technology, most existing text interventions for adolescent smoking cessation are not conversational. For example, the National Cancer Institute's SmokefreeTxt for Teens\footnote{SmokefreeText for Teens: https://teen.smokefree.gov/become-smokefree/smokefreeteen-signup} is limited to keywords such as ``yes,'' ``no,'' ``quit,'' or ``crave.'' One example of a conversational text-based intervention designed for adults is Bella~\footnote{Quit with Bella: https://www.solutions4health.co.uk/our-services/quitwithbella/}. Bella is an AI-powered chatbot for smoking cessation, developed primarily based on expert guidance, and was released in January 2018 in the UK. Bella is advertised as a ``coach,'' and the developers did not specify a theoretical orientation such as CBT. Data regarding efficacy for Bella has yet to be published, and it does not appear to be tailored to adolescents. 
We also surveyed the works that were developed using the MI/CBT framework. Researchers have been using natural language processing methods, such as recursive neural networks~\cite{tanana2015recursive,xiao2016behavioral} or attention networks~\cite{gibson2017attention}, to automatically code the transcriptions of motivational interviews~\cite{tanana2016comparison}. Tanana {\em et al.} developed ClientBot, a patient-like conversational agent, to train basic counseling skills~\cite{tanana2019development}. However, while these prior works are inspiring, none of them created a working chatbot that uses MI/CBT to promote smoking cessation. More recently, Almusharraf created an FSM-based chatbot that motivates people to quit smoking, following the counseling style of MI~\cite{almusharraf2019motivating}. This work, even though not accomplishing a full MI/CBT treatment, shed light on how such automated chatbots can be employed to help youth quit smoking.

\section{What's Next?}
Existing AI solutions in the smoking cessation field overpromise and underdeliver. Bella, the most advanced smoking cessation chatbot on the market, gets stuck when users input responses that are not prompted~\cite{szalacsi_2019}. For example, after the initial goal-setting session with Bella, if a user attempts to engage in general conversation with Bella, it continually responds ``I can't wait for your quit date,'' which likely discourages engagement. A significant innovation would be to really deliver a human-like chatbot experience by taking advances from AI and bringing them to the smoking cessation field. One potential direction is to use transcripts from therapy sessions to develop a smoking cessation chatbot for adolescents. A data-driven approach would allow for anticipation of the types of conversations that come up for adolescents during smoking cessation interventions, which should promote engagement.

\paragraph{Needs for Longer-Term Smoking Cessation Support}
While many studies have reported significant effects on smoking cessation treatments relative to control conditions, only one study reported significant intervention effects at 12-month follow-up~\cite{nih29}. Based on the limited evidence for the durability of treatment effects, a recent Cochrane review indicated that ``there is limited evidence that either behavioral support or smoking cessation medication increases the proportion of young people that stop smoking in the long-term''~\cite{nih30}. Adolescents may need smoking cessation support beyond the traditional three-to-four-month treatment model. While it may not be feasible to provide ongoing, in-person support for smoking cessation due to cost and the limited availability of providers, chatbots may be easily scaled for extended use to support long-term smoking cessation.

\paragraph{Role of Future Smoking Cessation Chatbots}
While chatbots may simulate human interaction, it is unlikely that they would replace mental health providers. Given the novelty of using chatbots for smoking cessation, the optimal role of chatbots in human mental healthcare provider work has yet to be determined. Ideally, chatbots would systematically deliver routine components of smoking cessation interventions and thus allow more time for providers to help with more complex aspects of smoking cessation. If developed as FDA-approved digital therapeutics, smoking cessation chatbots might be prescribed by providers in a manner similar to drugs, and providers might schedule follow-up appointments to review patient response to the treatment. Future work should explore which approaches to integrating chatbots in clinical care are optimal for smoking cessation.

\bibliographystyle{aaai}
\bibliography{aaai.bib}

\begin{thebibliography}{}

\bibitem[\protect\citeauthoryear{Almusharraf}{2019}]{almusharraf2019motivating}
Almusharraf, F.
\newblock 2019.
\newblock {\em Motivating Smokers to Quit Through a Computer-Based
  Conversational System}.
\newblock Ph.D. Dissertation.

\bibitem[\protect\citeauthoryear{Bachman \bgroup et al\mbox.\egroup
  }{2011}]{nih3}
Bachman, J.~G.; O'Malley, P.~M.; Johnston, L.~D.; Schulenberg, J.~E.; and
  Wallace~Jr, J.~M.
\newblock 2011.
\newblock Racial/ethnic differences in the relationship between parental
  education and substance use among us 8th-, 10th-, and 12th-grade students:
  findings from the monitoring the future project.
\newblock {\em Journal of studies on alcohol and drugs} 72(2):279--285.

\bibitem[\protect\citeauthoryear{CDC}{2006}]{nih6}
CDC.
\newblock 2006.
\newblock Use of cessation methods among smokers aged 16-24 years--united
  states, 2003.
\newblock {\em MMWR. Morbidity and mortality weekly report} 55(50):1351.

\bibitem[\protect\citeauthoryear{CDC}{2011}]{nih25}
CDC.
\newblock 2011.
\newblock Vital signs: current cigarette smoking among adults aged >= 18
  years--united states, 2005-2010.
\newblock {\em MMWR. Morbidity and mortality weekly report} 60(35):1207.

\bibitem[\protect\citeauthoryear{Fanshawe \bgroup et al\mbox.\egroup
  }{2017}]{nih30}
Fanshawe, T.~R.; Halliwell, W.; Lindson, N.; Aveyard, P.; Livingstone-Banks,
  J.; and Hartmann-Boyce, J.
\newblock 2017.
\newblock Tobacco cessation interventions for young people.
\newblock {\em Cochrane Database of Systematic Reviews} (11).

\bibitem[\protect\citeauthoryear{Fitzpatrick, Darcy, and
  Vierhile}{2017}]{nih33}
Fitzpatrick, K.~K.; Darcy, A.; and Vierhile, M.
\newblock 2017.
\newblock Delivering cognitive behavior therapy to young adults with symptoms
  of depression and anxiety using a fully automated conversational agent
  (woebot): a randomized controlled trial.
\newblock {\em JMIR mental health} 4(2):e19.

\bibitem[\protect\citeauthoryear{Gibson \bgroup et al\mbox.\egroup
  }{2017}]{gibson2017attention}
Gibson, J.; Can, D.; Georgiou, P.~G.; Atkins, D.~C.; and Narayanan, S.~S.
\newblock 2017.
\newblock Attention networks for modeling behaviors in addiction counseling.
\newblock In {\em INTERSPEECH},  3251--3255.

\bibitem[\protect\citeauthoryear{HHS}{2012}]{nih1}
HHS.
\newblock 2012.
\newblock Preventing tobacco use among youth and young adults: a report of the
  surgeon general.

\bibitem[\protect\citeauthoryear{Joffe \bgroup et al\mbox.\egroup
  }{2009}]{nih27}
Joffe, A.; McNeely, C.; Colantuoni, E.; An, M.-W.; Wang, W.; and Scharfstein,
  D.
\newblock 2009.
\newblock Evaluation of school-based smoking-cessation interventions for
  self-described adolescent smokers.
\newblock {\em Pediatrics} 124(2):e187--e194.

\bibitem[\protect\citeauthoryear{Johnston \bgroup et al\mbox.\egroup
  }{2015}]{nih4}
Johnston, L.~D.; O'Malley, P.~M.; Miech, R.~A.; Bachman, J.~G.; and
  Schulenberg, J.~E.
\newblock 2015.
\newblock Demographic subgroup trends among adolescents in the use of various
  licit and illicit drugs, 1975-2014. monitoring the future occasional paper
  series. paper 83.
\newblock {\em Institute for Social Research}.

\bibitem[\protect\citeauthoryear{Kong \bgroup et al\mbox.\egroup }{2017}]{nih9}
Kong, G.; Goldberg, A.~L.; Dallery, J.; and Krishnan-Sarin, S.
\newblock 2017.
\newblock An open-label pilot study of an intervention using mobile phones to
  deliver contingency management of tobacco abstinence to high school students.
\newblock {\em Experimental and clinical psychopharmacology} 25(5):333.

\bibitem[\protect\citeauthoryear{Lenhart \bgroup et al\mbox.\egroup
  }{2010}]{nih8}
Lenhart, A.; Ling, R.; Campbell, S.; and Purcell, K.
\newblock 2010.
\newblock Teens and mobile phones: Text messaging explodes as teens embrace it
  as the centerpiece of their communication strategies with friends.
\newblock {\em Pew Internet \& American Life Project}.

\bibitem[\protect\citeauthoryear{McClure \bgroup et al\mbox.\egroup
  }{2017}]{nih10}
McClure, E.~A.; Baker, N.~L.; Carpenter, M.~J.; Treiber, F.~A.; and Gray, K.~M.
\newblock 2017.
\newblock Attitudes and interest in technology-based treatment and the remote
  monitoring of smoking among adolescents and emerging adults.
\newblock {\em Journal of smoking cessation} 12(2):88--98.

\bibitem[\protect\citeauthoryear{Minary \bgroup et al\mbox.\egroup
  }{2013}]{nih29}
Minary, L.; Cambon, L.; Martini, H.; Wirth, N.; Acouetey, D.~S.; Thouvenot, F.;
  Maire, C.; Martinet, Y.; Bohadana, A.; Zmirou-Navier, D.; et~al.
\newblock 2013.
\newblock Efficacy of a smoking cessation program in a population of adolescent
  smokers in vocational schools: a public health evaluative controlled study.
\newblock {\em BMC public health} 13(1):149.

\bibitem[\protect\citeauthoryear{Molteni}{2018}]{molteni_2018}
Molteni, M.
\newblock 2018.
\newblock The chatbot therapist will see you now.

\bibitem[\protect\citeauthoryear{Peterson \bgroup et al\mbox.\egroup
  }{2009}]{nih28}
Peterson, A.~V.; Kealey, K.~A.; Mann, S.~L.; Marek, P.~M.; Ludman, E.~J.; Liu,
  J.; and Bricker, J.~B.
\newblock 2009.
\newblock Group-randomized trial of a proactive, personalized telephone
  counseling intervention for adolescent smoking cessation.
\newblock {\em JNCI: Journal of the National Cancer Institute}
  101(20):1378--1392.

\bibitem[\protect\citeauthoryear{Simon \bgroup et al\mbox.\egroup
  }{2015}]{simon2015socioeconomic}
Simon, P.; Camenga, D.~R.; Kong, G.; Cavallo, D.~A.; Connell, C.; Gutierrez,
  K.~M.; and Krishnan-Sarin, S.
\newblock 2015.
\newblock Socioeconomic disparities in electronic cigarette use among
  adolescents.
\newblock {\em Drug and Alcohol Dependence} 100(156):e205.

\bibitem[\protect\citeauthoryear{Singh}{2016}]{nih26}
Singh, T.
\newblock 2016.
\newblock Tobacco use among middle and high school students—united states,
  2011--2015.
\newblock {\em MMWR. Morbidity and mortality weekly report} 65.

\bibitem[\protect\citeauthoryear{Smith and Page}{2015}]{nih31}
Smith, A., and Page, D.
\newblock 2015.
\newblock Us smartphone use in 2015.
\newblock {\em Pew Research Center} 1.

\bibitem[\protect\citeauthoryear{Szalacsi}{2019}]{szalacsi_2019}
Szalacsi, B.
\newblock 2019.
\newblock Ai and quitting addictions: how i quit the "quit with bella" app and
  stayed a smoker.

\bibitem[\protect\citeauthoryear{Tanana \bgroup et al\mbox.\egroup
  }{2015}]{tanana2015recursive}
Tanana, M.; Hallgren, K.; Imel, Z.; Atkins, D.; Smyth, P.; and Srikumar, V.
\newblock 2015.
\newblock Recursive neural networks for coding therapist and patient behavior
  in motivational interviewing.
\newblock In {\em Proceedings of the 2nd Workshop on Computational Linguistics
  and Clinical Psychology: From Linguistic Signal to Clinical Reality},
  71--79.

\bibitem[\protect\citeauthoryear{Tanana \bgroup et al\mbox.\egroup
  }{2016}]{tanana2016comparison}
Tanana, M.; Hallgren, K.~A.; Imel, Z.~E.; Atkins, D.~C.; and Srikumar, V.
\newblock 2016.
\newblock A comparison of natural language processing methods for automated
  coding of motivational interviewing.
\newblock {\em Journal of substance abuse treatment} 65:43--50.

\bibitem[\protect\citeauthoryear{Tanana \bgroup et al\mbox.\egroup
  }{2019}]{tanana2019development}
Tanana, M.~J.; Soma, C.~S.; Srikumar, V.; Atkins, D.~C.; and Imel, Z.~E.
\newblock 2019.
\newblock Development and evaluation of clientbot: Patient-like conversational
  agent to train basic counseling skills.
\newblock {\em Journal of medical Internet research} 21(7):e12529.

\bibitem[\protect\citeauthoryear{Tworek \bgroup et al\mbox.\egroup
  }{2014}]{nih5}
Tworek, C.; Schauer, G.~L.; Wu, C.~C.; Malarcher, A.~M.; Jackson, K.~J.; and
  Hoffman, A.~C.
\newblock 2014.
\newblock Youth tobacco cessation: quitting intentions and past-year quit
  attempts.
\newblock {\em American journal of preventive medicine} 47(2):S15--S27.

\bibitem[\protect\citeauthoryear{Xiao \bgroup et al\mbox.\egroup
  }{2016}]{xiao2016behavioral}
Xiao, B.; Can, D.; Gibson, J.; Imel, Z.~E.; Atkins, D.~C.; Georgiou, P.~G.; and
  Narayanan, S.~S.
\newblock 2016.
\newblock Behavioral coding of therapist language in addiction counseling using
  recurrent neural networks.
\newblock In {\em Interspeech},  908--912.

\end{thebibliography}

\end{document}